\documentclass[twocolumn, prl, superscriptaddress,notitlepage]{revtex4-1}
\usepackage{graphicx}
\usepackage{physics}
\usepackage{epsfig}
\usepackage{color}
\usepackage{xspace}
\usepackage{array}
\usepackage{amsmath}
\usepackage{amsfonts}
\usepackage{ulem}
\usepackage[dvipsnames]{xcolor}
\usepackage{colortbl}
\usepackage[breaklinks,colorlinks,linkcolor=blue,citecolor=blue,urlcolor=blue]{hyperref}

\begin{document}
\title{Geometric Optimization of Quantum Control with Minimum Cost}

 \author{Chengming Tan}
 \affiliation{Hefei National Research Center for Physical Sciences at the Microscale and School of Physical Sciences, University of Science and Technology of China, Hefei 230026, China}
 \affiliation{Hefei National Laboratory, Hefei, 230088, China}

\author{Yuhao Cai}
\affiliation{MOE Key Laboratory for Nonequilibrium Synthesis and Modulation of Condensed Matter, Shaanxi Province Key Laboratory of Quantum
Information and Quantum Optoelectronic Devices, School of Physics, Xi'an Jiaotong University, Xi'an 710049, China}

 \author{Jinyi Zhang}
 \affiliation{Hefei National Research Center for Physical Sciences at the Microscale and School of Physical Sciences, University of Science and Technology of China, Hefei 230026, China}
 \affiliation{Hefei National Laboratory, Hefei, 230088, China}

\author{Shengli Ma}
\email{msl1987@xjtu.edu.cn}
\affiliation{MOE Key Laboratory for Nonequilibrium Synthesis and Modulation of Condensed Matter, Shaanxi Province Key Laboratory of Quantum
Information and Quantum Optoelectronic Devices, School of Physics, Xi'an Jiaotong University, Xi'an 710049, China}

\author{Chenwei Lv}
\email{lcw205046@gmail.com}
\affiliation{Homer L. Dodge Department of Physics and Astronomy, University of Oklahoma, Norman, OK, 73019, USA}

 \author{Ren Zhang}
 \email{renzhang@xjtu.edu.cn}
\affiliation{MOE Key Laboratory for Nonequilibrium Synthesis and Modulation of Condensed Matter, Shaanxi Province Key Laboratory of Quantum
Information and Quantum Optoelectronic Devices, School of Physics, Xi'an Jiaotong University, Xi'an 710049, China}
\affiliation{Hefei National Laboratory, Hefei, 230088, China}
\date{\today}

\begin{abstract}
We investigate the optimization of quantum control from a differential geometric perspective. In our approach, optimal control minimizes the cost associated with evolving a quantum state, with the cost quantified by the length of the trajectory on a relevant Riemannian manifold. We demonstrate the optimization protocol in systems with SU(2) and SU(1,1) dynamical symmetries, which encompass a broad range of physical systems. In these systems, the time evolution can be represented by trajectories on a three-dimensional manifold. Given the initial and final states, the minimum-cost quantum control corresponds to a geodesic on the manifold. When the trajectory between the initial and final states is specified, the minimum-cost control corresponds to a geodesic within a submanifold embedded in the three-dimensional space. This framework provides a geometric method for optimizing shortcuts to adiabatic driving.
\end{abstract}

\maketitle

{\it Introduction-.} Geometry plays a fundamental role in unifying different realms of physics. For example, fiber bundles provide the mathematical foundation for both gauge theory and general relativity. Introducing geometric concepts into physical problems usually leads to deeper and clearer insights.
Many important questions in quantum physics, such as the minimum time for state transfer~\cite{TO1, TO2, TO3, TO4, TO5, TO6, TO7, TO8, TO9, TO10, TO11, TO12, TO13, TO14}, optimal quantum circuit design~\cite{Geo_QSL, Geo1, CG, CG_Lv}, and the speed limits of quantum evolution~\cite{QSL1,QSL2,QSL3,QSl4,QSL5,QSL6,QSL7,QSL8,QSL9,QSL10}, can all be formulated as problems of finding geodesics on Riemannian manifolds. 

A central goal of quantum control is to achieve reliable operations with high efficiency, such as reaching the quantum speed limit while maintaining high fidelity~\cite{Geo2, FF1,FF2,FF3,FF4}.
Recent studies have revealed a significant trade-off between speed and energy cost, even in single-particle systems~\cite{Trade-off1,Trade-off2,Trade-off3}.
The energy cost has been modeled as the norm of the Hamiltonian~\cite{GTQD,GTQD2,GTQDE}. However, the anisotropicity of the energy cost is overlooked.
In quantum computing platforms, such as atoms arrays, superconducting circuits, and trapped ions~\cite{Atom,Atom1,Atom2,Atom3,Atom4,Atom5,Atom6,Ion,Ion1,Ion2,Ion3,Ion4,Ion5,SC,SC1,SC2,SC3,SC4}, operations are typically driven by lasers or electromagnetic fields with finite power and frequency range. These limitations make it crucial to design quantum control strategies that minimize the energy cost under constraints.

In this Letter, we present a quantum control strategy that minimizes the cost of the operation. We formulate the problem as finding geodesics in a Riemannian manifold or its submanifold, and the constraints determine their metrics.
Our main conclusions are as follows:  
(1) Given the initial and final states, we identify the optimal trajectory that drives the system with the minimal cost.  
(2) When the initial and final states, along with the intermediate states, are specified, we construct a time-dependent Hamiltonian that realizes the targeted evolution while minimizing the cost.  
(3) We demonstrate our approach in systems with SU(2) or SU(1,1) dynamical symmetries, covering a broad class of physical settings~\cite{BOOK1, su112, su113, su114, su115, su117, su118, su1111, su1112, su1116, su111, su116, su119, su1110, su1113, su1114, su1115}.

{\it SU(2) dynamical system-.}
We begin with the SU(2) dynamically symmetric system. 
Without loss of generality, we consider a spin-1/2 particle subject to a time-dependent magnetic field ${\bf B}(t) = (B_{x}(t), B_{y}(t), B_{z}(t))$. Since the Pauli matrices $\hat{\sigma}_{x,y,z}$ 
provide a two-dimensional representation of the ${\mathfrak {su}}(2)$ algebra, the states of an SU(2) dynamical system are represented by points on a two-dimensional manifold, namely the Bloch sphere.  The wavefunction is written as $|\varphi(\theta,\phi)\rangle = (-\text{e}^{-i\phi/2}\sin(\theta/2), \text{e}^{i\phi/2}\cos(\theta/2))^{\rm T}$, where $\theta$ and $\phi$ are the spherical coordinates on the Bloch sphere. A global phase factor $\text{e}^{i\eta}$ can also be attached to the wavefunction. As we will show, the global phase $\eta$ plays a crucial role in minimizing cost of the control. The time evolution of the system can thus be viewed as a trajectory in the parameter space $(\theta, \phi, \eta)$, where all parameters are time-dependent. For brevity, we omit the explicit time dependence in the following discussion.

To quantify the cost associated with a given trajectory, we parameterize the evolution operator as $\hat{U} = \exp(-i\frac{\phi}{2}\hat\sigma_z) \exp(-i\frac{\theta}{2}\hat\sigma_y) \exp(-i\eta\hat\sigma_z)$. It can be verified that ${\rm e}^{i\eta}|\varphi(\theta,\phi)\rangle = \hat{U}(\theta,\phi,\eta)|\varphi(0,0)\rangle$. 
Therefore, we can design various magnetic field configurations that drive the system from the initial to the final state along different paths parameterized by $\theta$, $\phi$, and $\eta$. A natural question arises: which trajectory minimizes the cost?
To answer this, we compute the time-dependent Hamiltonian 
\begin{equation}
\hat{H}(t) = i(\partial_t\hat{U}(t))\hat{U}^{-1}(t) = \sum_{i=x,y,z} B_i(t)\hat\sigma_i,
\label{Eq.su2H}
\end{equation}
where we set $\mu_B = 1$ and $\hbar = 1$. The components of the time-dependent magnetic field are obtained via $B_i(t) = \text{Tr}[\hat{H}(t)\hat\sigma_i]/2$. (For the explicit expressions, please refer to Supplemental Materials  (SM) \cite{SM}).
Thus, once the time dependence of $\theta(t)$, $\phi(t)$, and $\eta(t)$ is specified, the corresponding Hamiltonian is fully determined.

We define 
the cost function as
\begin{equation}
ds^2 = \sum_{i=x,y,z} {\cal I}_i^2 B_i^2 dt^2,
\label{Eq.CEle}
\end{equation}
where the dimensionless coefficient ${\cal I}_i$ represents the difficulty of applying the external field along the $i$-th direction~\cite{CG,CG_Lv}. 
The total cost of evolution is then given by the integral of the cost function along the trajectory
$s = \int_{s_{\rm i}}^{s_{\rm f}} ds$
Substituting the explicit expressions of $B_i(t)$ into Eq.~(\ref{Eq.CEle}), the cost function is recast as the line element in the parameter space,
\begin{equation}
\begin{aligned}
ds^2 =& g_{\eta\eta}d\eta^2+g_{\theta\theta}d\theta^2+g_{\phi\phi}d\phi^2\\
&+2g_{\eta\theta} d\eta d\theta +2g_{\eta\phi} d\eta d\phi,
\label{Eq.su2gM}
\end{aligned}
\end{equation}
which defines a metric of the parameter space $(\theta, \phi, \eta)$. The explicit expressions of metric is presented in \cite{SM}. 
Thus, the problem of optimizing the evolution with minimum cost is transformed into finding the shortest trajectory under the metric in Eq.~(\ref{Eq.su2gM}). Notice that $\eta$ is a global phase, wavefunctions differing by $\eta$ represent the same physical state. 
Therefore, we now consider two different scenarios of optimization.

\begin{figure}[t]
	\centering
	\includegraphics[width=0.45\textwidth]{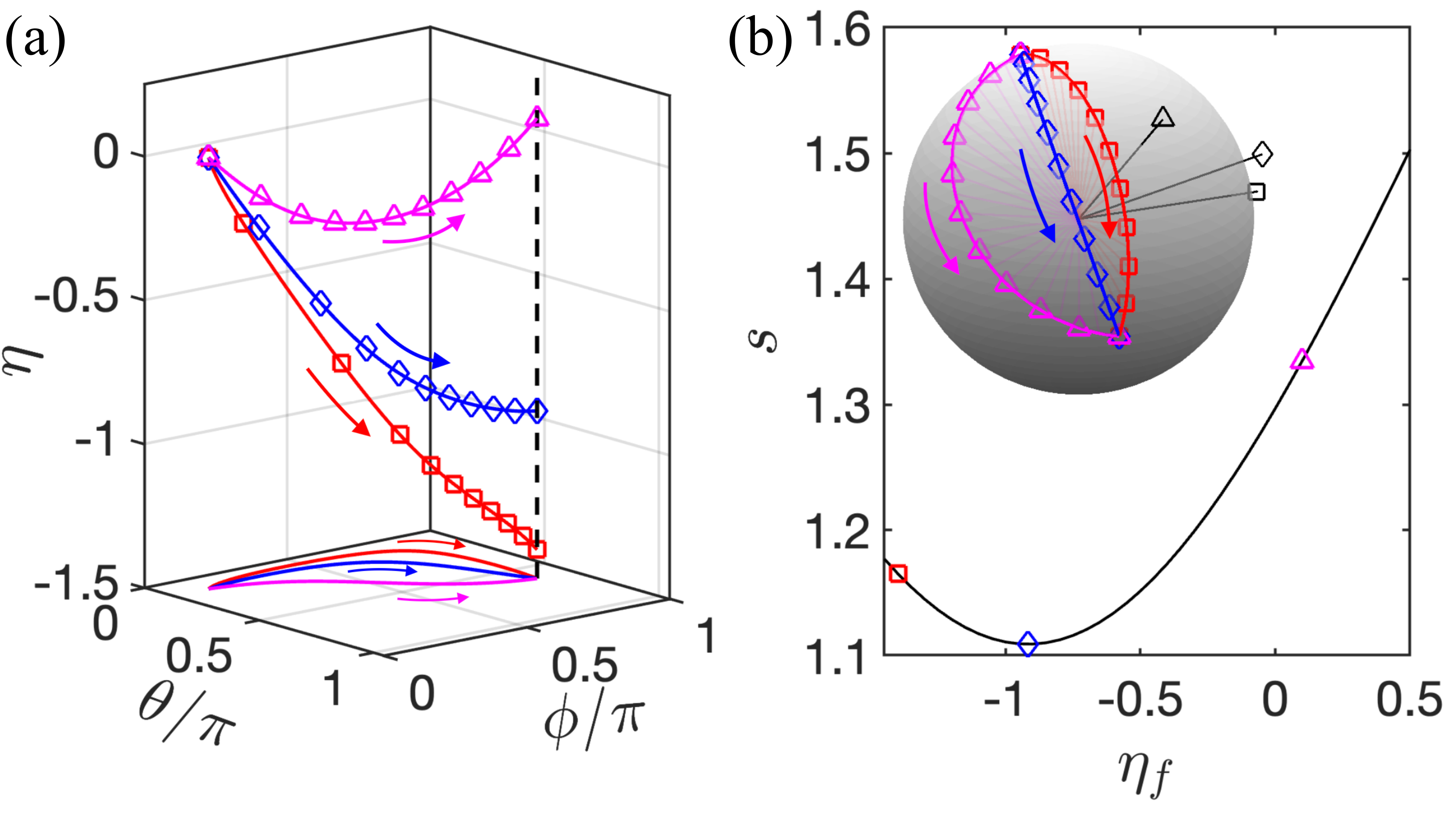}
	\caption{Geodesics from $(\theta_{\rm i}, \phi_{\rm i}, 0)$ to $(\theta_{\rm f}, \phi_{\rm f}, \eta_{\rm f})$ in the SU(2) system. Markers indicate different final phases $\eta_{\rm f}$. (a) Colored lines with markers: geodesics in $(\theta, \phi, \eta)$ space; lines without markers: projections onto $(\theta, \phi)$. The blue diamond line projects to a geodesic in the reduced space. Parameters: $\theta_{\rm i} = \pi/8$, $\theta_{\rm f} = 5\pi/8$, $\phi_{\rm i} = \pi/8$, $\phi_{\rm f} = 7\pi/8$. (b) Geodesic length vs. $\eta_{\rm f}$. Inset: projected trajectories and driving field directions on the Bloch sphere.}
	\label{Fig1}
\end{figure}

{ (I) Given initial and final states with fixed phases.}
Starting from a given initial state, we now determine the minimum cost required to reach a target state with fixed phases. 
Geometrically, this corresponds to finding the geodesic connecting two points in the parameter space $(\theta, \phi, \eta)$ with the metric given in Eq.~(\ref{Eq.su2gM}).
For simplicity, we consider the isotropic case with ${\cal I}_i = 1$, where the metric is rewritten as
\begin{equation}
ds^2 =\left(d\eta+\frac{1}{2}\cos\theta d\phi\right)^2+\frac{1}{4}(d\theta^2+\sin^2\theta d\phi^2).
\label{Eq.su2M}
\end{equation}
Then the geodesic equations are written as \cite{SM}
\begin{equation}
\begin{aligned}
\partial_t^2\eta &= -\cot \theta \, \partial_t\eta \, \partial_t\theta + \frac{1}{2}\csc\theta \, (\partial_t\phi \, \partial_t\theta),\\
\partial_t^2\theta &= -2\sin\theta \, (\partial_t\eta \, \partial_t\phi),\\
\partial_t^2\phi &= 2\csc\theta \, (\partial_t\eta \, \partial_t\theta) - \cot\theta \, (\partial_t\phi \, \partial_t\theta),
\end{aligned}
\label{Eq.su2G}
\end{equation}
where the initial point $(\theta_{\rm i}, \phi_{\rm i}, \eta_{\rm i})$ and target point $(\theta_{\rm f}, \phi_{\rm f}, \eta_{\rm f})$ serve as the boundary conditions. 
By solving Eq.~(\ref{Eq.su2G}), we obtain the geodesic trajectory parameterized by $(\theta(t), \phi(t), \eta(t))$. Physically, this corresponds to finding the instantaneous magnetic field $(B_x(t), B_y(t), B_z(t))$ that drives the system along the minimum-cost path. 
In Fig.~\ref{Fig1}(a), we illustrate the geodesics between different points, shown as colored lines with markers. The parameters are chosen as $\theta_{\rm i} = \pi/8$, $\theta_{\rm f} = 5\pi/8$, $\phi_{\rm i} = \pi/8$, $\phi_{\rm f} = 7\pi/8$, with $\eta_{\rm i} = 0$, and varying $\eta_{\rm f}$. Each trajectory shown in Fig.~\ref{Fig1}(a) represents the shortest path corresponding to a different final global phase $\eta_{\rm f}$.

Since different choices of $\eta_{\rm f}$ correspond to the same physical state, the cost can be further reduced by selecting an optimal final phase. Geometrically, this corresponds to finding the shortest geodesic from the initial point $(\theta_{\rm i}, \phi_{\rm i}, 0)$ to the “fiber” (black dashed line in Fig.~\ref{Fig1}(a)) over $(\theta_{\rm f}, \phi_{\rm f})$.
According to the metric in Eq.~(\ref{Eq.su2M}), minimizing the cost requires eliminating the contribution from the first term. Therefore, the optimal phase $\eta(t)$ along the trajectory satisfies
\begin{equation}
\eta_{\rm opt}(t) = -\frac{1}{2} \int_{t_{\rm i}}^{t} \cos\theta(\tau) \left[\partial_\tau\phi(\tau)\right] d\tau.
\label{Eq.su2MC}
\end{equation}
Under this condition, the metric reduces to
$ds^{2} = \frac{1}{4} \left(d\theta^2 + \sin^2\theta d\phi^2\right)$,
which describes a two-sphere with radius $1/2$.
The geodesic equations on this reduced manifold are
\begin{equation}
\begin{aligned}
\partial_t^2\theta &= \sin\theta \cos\theta \left(\partial_t\phi\right)^2,\\
\partial_t^2\phi &= -2\cot\theta \left(\partial_t\phi \, \partial_t\theta\right),
\end{aligned}
\label{GoedesicS2}
\end{equation}
with boundary conditions $(\theta_{\rm i}, \phi_{\rm i})$ and $(\theta_{\rm f}, \phi_{\rm f})$.
Solving Eq.~(\ref{GoedesicS2}) yields the geodesic $(\theta(t), \phi(t))$ along the minor arc on the Bloch sphere. By combining this trajectory with the optimal phase $\eta_{\rm opt}(t)$ from Eq.~(\ref{Eq.su2MC}), we obtain the shortest path connecting $(\theta_{\rm i}, \phi_{\rm i}, 0)$ to the fiber over $(\theta_{\rm f}, \phi_{\rm f})$ in the three-dimensional manifold. The corresponding Hamiltonian, constructed from this trajectory, realizes the minimum-cost evolution.


\begin{figure}[t]
	\centering
	\includegraphics[width=0.45\textwidth]{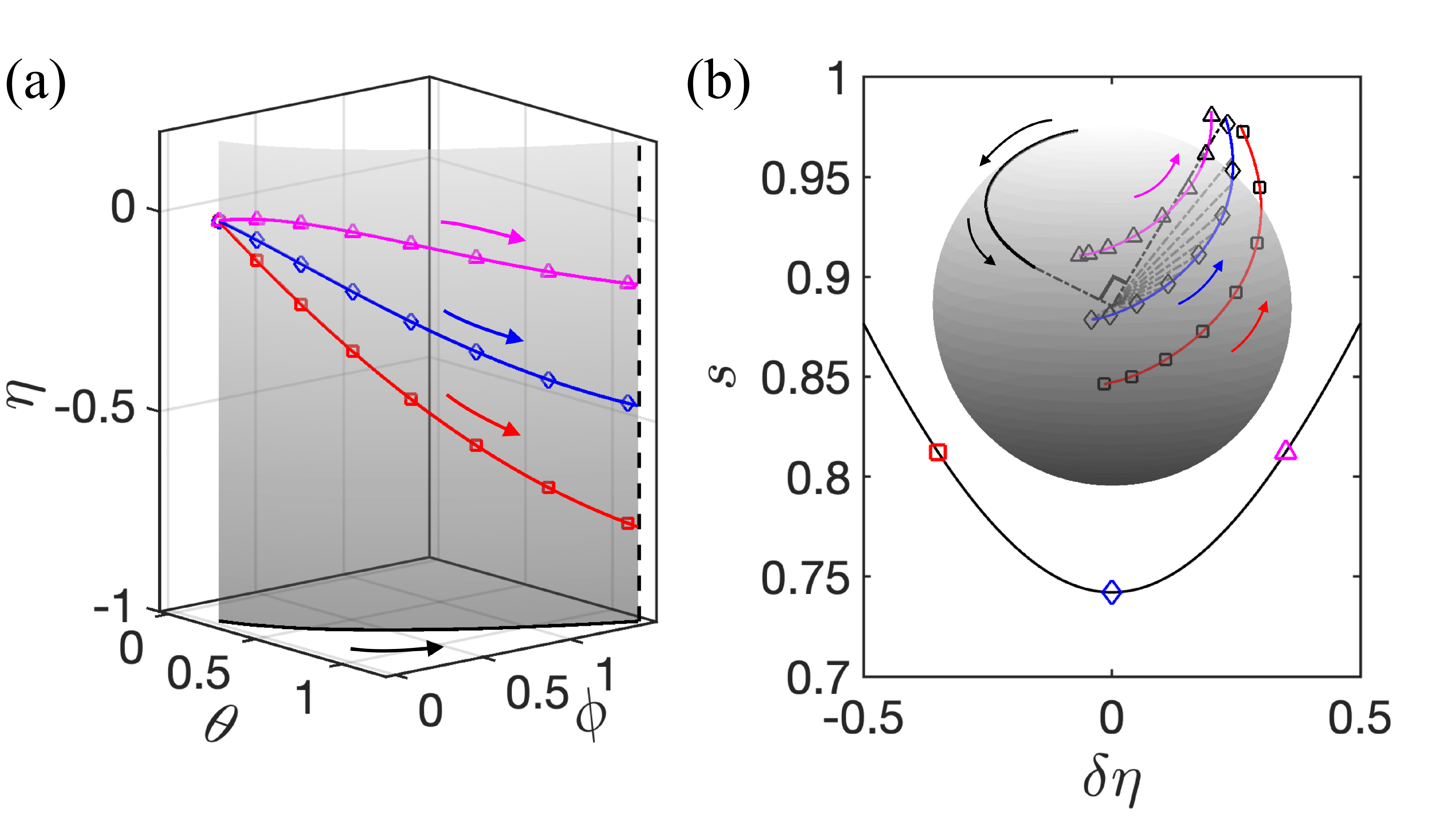}
	\caption{Evolution along a fixed trajectory $(\theta(t), \phi(t))$ in the SU(2) system. Markers indicate different deviations $\delta\eta$ from the optimal phase $\eta_{\rm opt}(t)$ in Eq.~(\ref{Eq.su2MC}). (a) Colored lines are trajectories with varying $\delta\eta$; all project onto the same path (black solid line). The blue diamond line minimizes the cost. (b) Trajectory lengths vs. $\delta\eta$. Inset: state projection (black line) and driving fields (black markers) on the Bloch sphere.}
	\label{Fig2}
\end{figure}


Fig.~\ref{Fig1}(a) 
 illustrates the shortest geodesic, shown as a blue line with diamond markers. Its projection remains a geodesic on the reduced two-dimensional manifold. In contrast, other trajectories, indicated by red rectangles and pink triangles, do not maintain this property.
Fig.~\ref{Fig1}(b) displays the geodesic lengths as a function of the final global phase $\eta_{\rm f}$. It is evident that the length increases as $\eta_{\rm f}$ deviates from the optimal value. The inset of Fig.~\ref{Fig1}(b) shows the corresponding state trajectories projected onto the Bloch sphere. Here, black markers indicate the direction of the magnetic field $B_i$, matched to the shapes of the state trajectories.
For the minimum-cost driving, the state trajectory forms a minor arc on the Bloch sphere, with the magnetic field perpendicular to the plane spanned by the arc and the radius. 

{ (II). Given a specified trajectory.}
In this case, in addition to specifying the initial and target states $(\theta_{\rm i}, \phi_{\rm i})$ and $(\theta_{\rm f}, \phi_{\rm f})$ on the Bloch sphere, we further constrain the evolution by a designated trajectory between them, given by $(\theta(t), \phi(t))$ for $t \in [0, t_{\rm f}]$ with $t_{\rm i} = 0$.
One strategy to drive the system along the prescribed path is adiabatic evolution, where the magnetic field remains aligned with $(\theta(t), \phi(t))$. However, in the adiabatic limit, the accumulated phase $d\eta$, and hence the cost as defined in our framework, diverges. 
An alternative is to employ shortcut-to-adiabatic-driving (STAD) protocols, which speed up the state transfer while suppressing unwanted excitations~\cite{Trade-off1, SA1, SA2, SA4, SA5, SA6, SA7, GTQD2, GTQD}. A central question then arises: Is there an optimal STAD protocol that minimizes the cost along a given trajectory?

Similar to (I), the global phase $\eta$ provides additional flexibility to minimize the cost defined by Eq.~(\ref{Eq.su2gM}).
We illustrate this with a specific example.
The solid black line in Fig.~\ref{Fig2}(a) is 
given by
\begin{equation}
\label{su2-exam}
\theta(t) = t/t_f+1/5;\ \phi(t) = \theta^2(t),
\end{equation}
for $t\in[0, t_{\rm f}]$. 
By attaching a fiber to each point along the trajectory, we define a submanifold (indicated by the gray surface) embedded in the full parameter space. Any trajectory within this submanifold that connects $(\theta_{\rm i}, \phi_{\rm i}, 0)$ to the fiber over $(\theta_{\rm f}, \phi_{\rm f})$ represents a possible STAD protocol, as shown by the colored lines with markers.
Without loss of generality, we still consider the isotropic case with ${\cal I}_i=1$.
Substituting Eq.~(\ref{su2-exam}) into Eq.~(\ref{Eq.su2MC}) yields the shortest trajectory, denoted by the blue line with diamond markers in Fig.~\ref{Fig2}(a).
To confirm the optimality of this trajectory, we introduce perturbations by adding $\delta\eta\sin(\pi t/3t_{\rm f})$ to $\eta(t)$. The resulting trajectories 
are shown by the rectangle and triangle markers in Fig.~\ref{Fig2}(a), with their associated costs displayed in Fig.~\ref{Fig2}(b). Clearly, deviation from the optimal $\eta(t)$ leads to an increased cost.
For each trajectory in the submanifold, the instantaneous magnetic field ${\bf B}(t)$ can be obtained, as shown by the colored lines with markers in the inset of Fig.~\ref{Fig2}(b). We find that, for the minimum-cost STAD, the instantaneous magnetic field is always perpendicular to the state trajectory (the black curve) on the Bloch sphere. This perpendicularity condition may not generally hold for other, non-optimal protocols.


We now highlight the geometric origin underlying the 
optimal STAD. Substituting Eq.~(\ref{su2-exam}) into Eq.~(\ref{Eq.su2M}) yields the induced metric on the submanifold spanned by $(t, \eta)$. The problem then reduces to finding the shortest geodesic connecting the initial point $(t=0, \eta=0)$ to the fiber at $t = t_{\rm f}$ on the submanifold \cite{SM}. By solving the corresponding geodesic equation, we verify that the blue trajectory shown in Fig.~\ref{Fig2}(a), which minimizes the cost, indeed corresponds to the shortest geodesic within the submanifold.
Thus, the minimum-cost STAD can be understood as the geodesic motion on the induced two-dimensional manifold in the parameter space.

{\it SU(1,1) dynamical system-.}
We now turn to systems with SU(1,1) dynamical symmetry. The generators of the SU(1,1) group satisfy the commutation relations
$[\hat{K}_0, \hat{K}_1] = i\hat{K}_2, [\hat{K}_1, \hat{K}_2] = -i\hat{K}_0, [\hat{K}_2, \hat{K}_0] = i\hat{K}_1$.
Representations of the ${\mathfrak{su}}(1,1)$ algebra appear widely across various areas of physics, including harmonic and inverted harmonic oscillators, single- and two-mode squeezing operations, and spin-1/2 particles in complex magnetic fields, among others~\cite{su112, su113, su114, su115, su117, su118, su1111, su1112, su1116}.

Analogous to the SU(2) case, a quantum state $|\psi(\rho, \phi)\rangle$ in an SU(1,1) dynamically symmetric system can be represented by a point on a two-dimensional manifold parameterized by $\rho$ and $\phi$. To quantify the cost of transporting the state from $|\psi(0, 0)\rangle$ to $|\psi(\rho, \phi)\rangle$, we parameterize the evolution operator as
$\hat{U} = \exp\left(-i{\phi}\hat{K}_0\right)\exp\left(-i{\rho}\hat{K}_2\right)\exp\left(-i2\eta\hat{K}_0\right)$,
where $\rho \in [0, +\infty)$. 
We adopt the representation of a spin-1/2 particle in complex magnetic fields, where $\hat{K}_0=\hat\sigma_z/2$, $\hat{K}_1=i\hat\sigma_x/2$ and $\hat{K}_2=i\hat\sigma_y/2$.
Therefore, $|\psi(\rho, \phi)\rangle =  (-\text{e}^{-i\phi/2}\sinh(\rho/2), \text{e}^{i\phi/2}\cosh(\rho/2))^{\rm T}$.
It can be verified that ${\rm e}^{i\eta}|\psi(\rho, \phi)\rangle = \hat{U} |\psi(0, 0)\rangle$. 
The evolution, generated by the time-dependent Hamiltonian $\hat{H}(t) = i\left(\partial_t\hat{U}(t)\right)\hat{U}^{-1}(t) = \sum_i \xi_i(t) \hat{K}_i$, traces a trajectory in the parameter space $(\rho, \phi, \eta)$.
The field strengths $\xi_i(t)$ are extracted from the Hamiltonian via $\xi_i(t) = 2\text{Tr}[\hat{H}(t)\hat{K}_i^\dagger]$ \cite{SM}. 
For the isotropic case, the cost function is defined as $ds^2 = \frac{1}{4}(\xi_0^2 + \xi_1^2 + \xi_2^2) dt^2$.
Substituting the expressions for $\xi_i(t)$ yields the induced metric of the parameter space
\begin{equation}
\label{Eq.su11M}
\begin{aligned}
ds^2 =& \left(\sqrt{\cosh2\rho}d\eta+\frac{\cosh\rho}{2\sqrt{\cosh2\rho}} d\phi\right)^2\\
&+\frac{1}{4}\left(d\rho^2+\frac{\sinh^2\rho}{\cosh(2\rho)} d\phi^2\right).
\end{aligned}
\end{equation}
With this metric, optimal quantum control in the SU(1,1) system can be engineered following a procedure analogous to the SU(2) case. 
In the following, we explore two distinct scenarios as well.

\begin{figure}[t]
	\centering
	\includegraphics[width=0.45\textwidth]{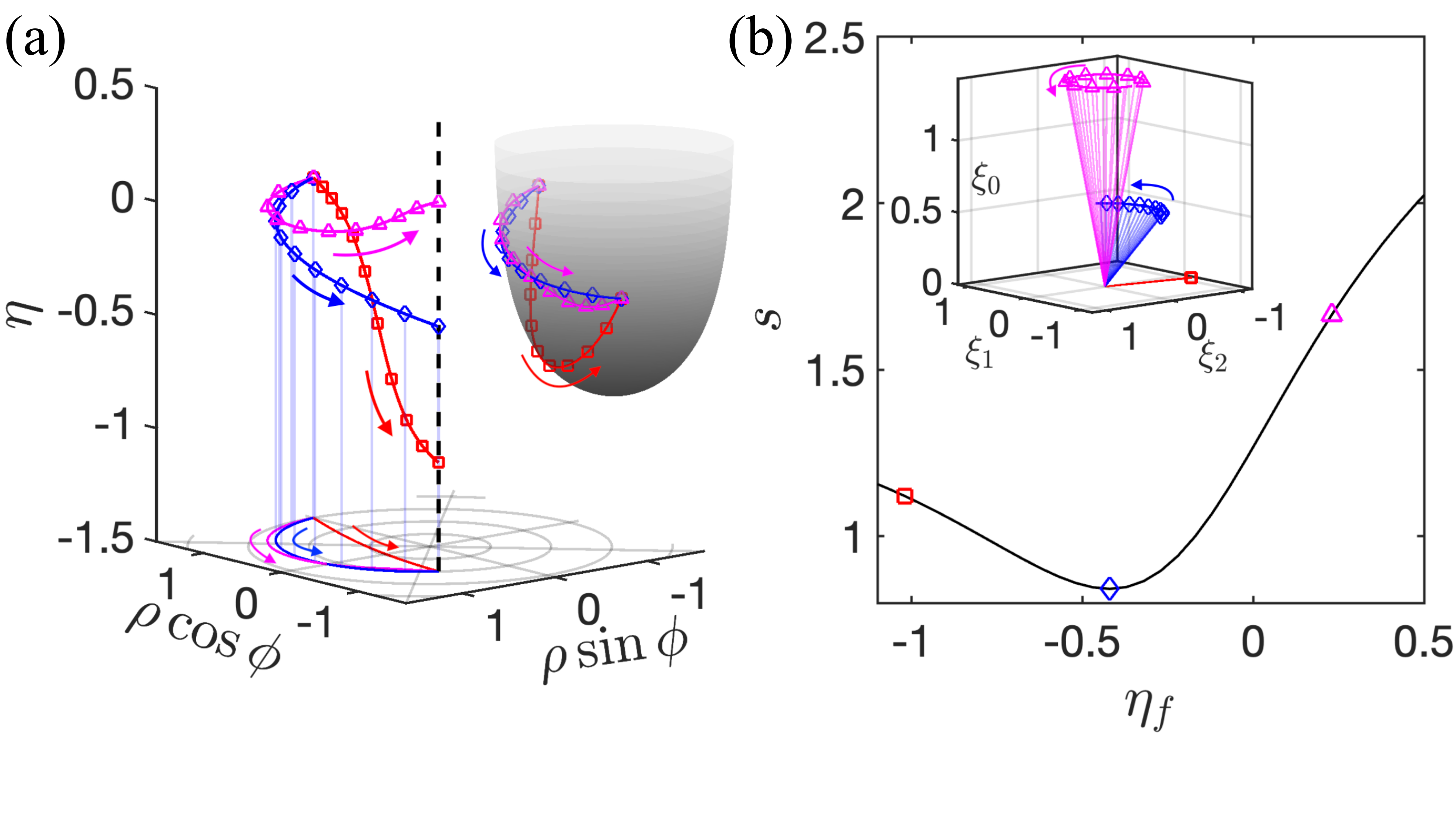}
	\caption{Geodesics from $(\rho_{\rm i}, \phi_{\rm i}, 0)$ to $(\rho_{\rm f}, \phi_{\rm f}, \eta_{\rm f})$ in the SU(1,1) systems. Markers indicate different final phases $\eta_{\rm f}$. (a) Colored lines with markers: geodesics in $(\rho, \phi, \eta)$ space; lines without markers: projections onto the $(\rho, \phi)$ plane. The blue diamond line projects to a geodesic in the reduced space. Parameters: $\rho_{\rm i} = 1.5$, $\phi_{\rm i} = 0$, $\rho_{\rm f} = 1.0$, $\phi_{\rm f} = 4\pi/5$. Inset: reduced curved surface embedded in flat 3D space. (b) Geodesic length vs. $\eta_{\rm f}$. Inset: corresponding driving fields.}
	\label{Fig3}
\end{figure}

{ (I) Given initial and final states with fixed global phase.}
We represent the the initial and final states by $(\rho_{\rm i}, \phi_{\rm i}, 0)$ and $(\rho_{\rm f}, \phi_{\rm f}, \eta_{\rm f})$, respectively, and determine the minimum-cost trajectory by solving the geodesic equations associated with the metric in Eq.~(\ref{Eq.su11M}) \cite{SM},
\begin{equation}
\begin{aligned}
\partial_t^2\eta &= -3\coth\rho\, (\partial_t\eta\, \partial_t\rho) - \frac{1}{2}\csch\rho\, (\partial_t\rho\, \partial_t\phi), \\
\partial_t^2\rho &= 4\sinh(2\rho)(\partial_t\eta)^2 + 2\sinh\rho\, (\partial_t\eta\, \partial_t\phi), \\
\partial_t^2\phi &= \frac{4 + 2\cosh(2\rho)}{\sinh\rho}(\partial_t\eta\, \partial_t\rho) + \coth\rho\, (\partial_t\rho\, \partial_t\phi).
\end{aligned}
\end{equation}
In Fig.~\ref{Fig3}(a), colored lines with markers represent geodesics connecting the initial state $|\psi(\rho_{\rm i}, \phi_{\rm i})\rangle$ to final states $e^{i\eta_{\rm f}}|\psi(\rho_{\rm f}, \phi_{\rm f})\rangle$ for different values of $\eta_{\rm f}$. Each trajectory corresponds to a minimum-cost evolution under the constraint of a fixed final global phase.

The cost can be further reduced by selecting an optimal $\eta_{\rm f}$. 
According to Eq.~(\ref{Eq.su11M}), the optimal phase trajectory is given by
\begin{equation}
\eta_{\rm opt}(t) = -\int_{t_{\rm i}}^{t} \frac{\cosh\rho(\tau)}{2\cosh2\rho(\tau)} \left[\partial_\tau\phi(\tau)\right] d\tau,
\label{Eq.su11rc}
\end{equation}
and the reduced metric becomes $ds^2 = (d\rho^2+\sinh^2\rho/\cosh2\rho d\phi^2)/4$.
The optimal path on the reduced manifold is obtained by solving the geodesic equations
\begin{equation}
\label{reduceGeodsu11}
\begin{aligned}
\partial_t^2\rho &= \text{sech}(2\rho)\tanh(2\rho)\left(\partial_t\phi\right)^2/2, \\
\partial_t^2\phi &= -2\coth\rho\, \text{sech}(2\rho) \, \partial_t\rho\, \partial_t\phi,
\end{aligned}
\end{equation}
with boundary conditions $(\rho_{\rm i}, \phi_{\rm i})$ and $(\rho_{\rm f}, \phi_{\rm f})$.
Solving Eq.~(\ref{reduceGeodsu11}) yields the optimal trajectory $(\rho(t), \phi(t))$, which is then substituted into Eq.~(\ref{Eq.su11rc}) to determine $\eta_{\rm opt}(t)$. The resulting geodesic $(\rho(t), \phi(t), \eta_{\rm opt}(t))$ for $t \in [t_{\rm i}, t_{\rm f}]$ is shown as the blue line with diamond in Fig.~\ref{Fig3}(a). Its projection in the $\rho$–$\phi$ plane corresponds to the solution of Eq.~(\ref{reduceGeodsu11}).
Figure~\ref{Fig3}(b) shows the geodesic lengths for different values of $\eta_{\rm f}$, confirming that the path constructed from Eqs.~(\ref{Eq.su11rc}) and~(\ref{reduceGeodsu11}) indeed achieves the minimum cost. The corresponding driving field $(\xi_0(t), \xi_1(t), \xi_2(t))$ is illustrated in the inset of Fig. \ref{Fig3}(a).

\begin{figure}[t]
	\centering
	\includegraphics[width=0.45\textwidth]{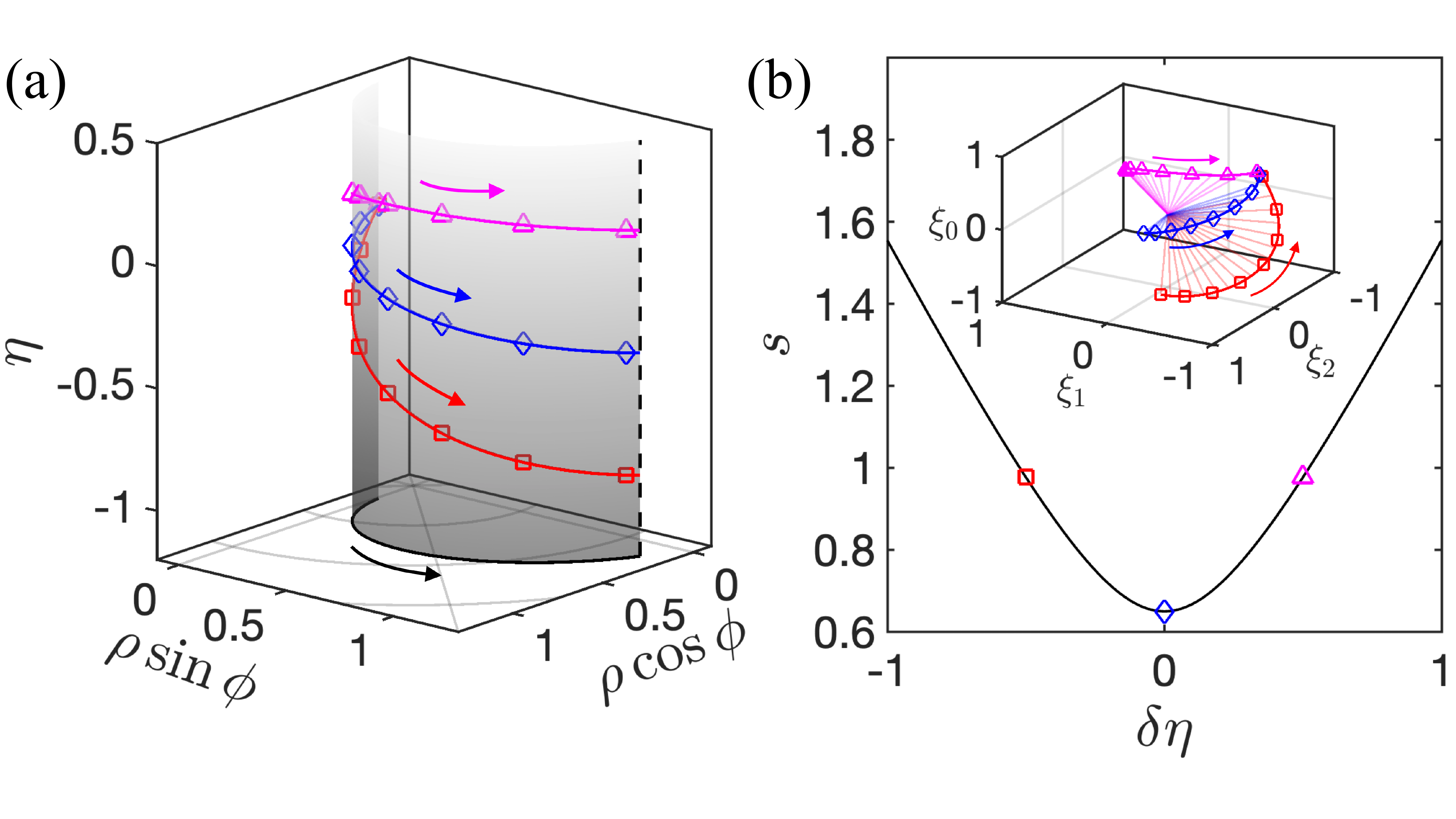}
	\caption{Evolution along a fixed trajectory $(\rho(t), \phi(t))$ in the SU(1,1) systems. Markers indicate deviations $\delta\eta$ from the optimal phase $\eta_{\rm opt}(t)$. (a) Colored lines are trajectories with varying $\delta \eta$; all project onto the same path (black solid line). The blue diamond line corresponds to the minimum-cost evolution. (b) Trajectory length vs.~$\delta\eta$. Inset: corresponding driving fields.}
	\label{Fig4}
\end{figure}

To visualize the state evolution on the reduced $\rho-\phi$ manifold, we embed it into a three-dimensional flat Euclidean space as a curved surface. The resulting geometry, shown in the inset of Fig.~\ref{Fig3}(a), serves as an SU(1,1) counterpart of the Bloch sphere for SU(2) systems. The geodesic obtained from the optimal trajectory (blue line with diamond) remains a geodesic under this embedding, as illustrated by its projection onto the curved surface.

{ (II) Given a specified trajectory.}
Similar to the SU(2) case, we now consider scenarios in which the trajectory of the state evolution is fixed. As an example, we specify the trajectory
\begin{equation}
\rho(t) = t/t_f+1/5;\ \phi(t) = \rho^2(t),
\label{Eq.su11p}
\end{equation}
for $t \in [0, t_{\rm f}]$, shown as the black solid line in Fig.~\ref{Fig4}(a).
By attaching a fiber to each point along this trajectory, we obtain an embedded surface in the parameter space, represented by the gray surface in Fig.~\ref{Fig4}(a). This surface captures all possible STAD for the given trajectory $(\rho(t), \phi(t))$, while allowing freedom in the global phase $\eta(t)$. The final states thus lie along the fiber above $(\rho_{\rm f}, \phi_{\rm f})$, indicated by the black dashed line.

Substituting Eq.~(\ref{Eq.su11p}) into Eq.~(\ref{Eq.su11rc}) yields the optimal phase trajectory $\eta_{\rm opt}(t)$. Together with the specified $(\rho(t), \phi(t))$, this determines the shortest path on the embedded surface, shown as the blue line with diamond markers in Fig.~\ref{Fig4}(a). The corresponding optimal driving field $(\xi_0(t), \xi_1(t), \xi_2(t))$ can be derived analytically, or alternatively obtained by solving the geodesic equations on the embedded surface \cite{SM}.
To verify that $\eta_{\rm opt}(t)$ minimizes the cost, we introduce a small deviation of the form
$\eta(t) = \eta_{\rm opt}(t) + \delta\eta \sin\left(\frac{\pi t}{2t_{\rm f}}\right)$,
and evaluate the trajectory length for different values of $\delta\eta$. As shown in Fig.~\ref{Fig4}(b), the trajectory with $\delta\eta = 0$ (diamond marker) corresponds to the shortest length, while deviations from the optimal phase result in increased cost.
The corresponding driving fields for these trajectories are illustrated in the inset of Fig.~\ref{Fig4}(b), showing time-dependent variations in both field strength and direction.

{\it Conclusion-.} 
We have investigated the optimization of quantum control in dynamically symmetric systems through a geometric framework. The minimum-cost control corresponds to the shortest geodesic on a Riemannian manifold defined by the system's parameters. We demonstrated this approach using SU(2) and SU(1,1) dynamical symmetries.
When only the initial and final states are specified, the optimal trajectory is the geodesic on the full three-dimensional manifold. When the evolution path between the endpoints is prescribed, the optimal STAD lies on a two-dimensional embedded submanifold.
Our method is readily generalizable to systems with higher symmetries and 
anisotropic cost functions, as suggested by Eq.~(\ref{Eq.CEle}), opens avenues for further exploration in a broader class of quantum control problems.

\begin{acknowledgments}
We are grateful to Qi Zhou, Peng Zhang, Xiaoling Cui, Yangqian Yan and Liang Mao for helpful discussions.
This work is supported by the Innovation Program for Quantum Science and Technology (Grants No.2021ZD0302001), NSFC (Grants No.12174300 and No.12374248), the Fundamental Research Funds for the Central Universities (Grant No. 71211819000001) and Tang Scholar.
\end{acknowledgments}

%


\end{document}


\onecolumngrid
\setcounter{equation}{0}
\setcounter{figure}{0}
\setcounter{table}{0}
\makeatletter
\renewcommand{\theequation}{S\arabic{equation}}
\renewcommand{\thefigure}{S\arabic{figure}}
\renewcommand{\thetable}{S\arabic{table}}

\author{Chengming Tan}
 \affiliation{Hefei National Research Center for Physical Sciences at the Microscale and School of Physical Sciences, University of Science and Technology of China, Hefei 230026, China}
 \affiliation{Hefei National Laboratory, Hefei, 230088, China}

\author{Yuhao Cai}
\affiliation{MOE Key Laboratory for Nonequilibrium Synthesis and Modulation of Condensed Matter, Shaanxi Province Key Laboratory of Quantum
Information and Quantum Optoelectronic Devices, School of Physics, Xi'an Jiaotong University, Xi'an 710049, China}

 \author{Jinyi Zhang}
 \affiliation{Hefei National Research Center for Physical Sciences at the Microscale and School of Physical Sciences, University of Science and Technology of China, Hefei 230026, China}
 \affiliation{Hefei National Laboratory, Hefei, 230088, China}

\author{Shengli Ma}
 \email{msl1987@xjtu.edu.cn}
\affiliation{MOE Key Laboratory for Nonequilibrium Synthesis and Modulation of Condensed Matter, Shaanxi Province Key Laboratory of Quantum
Information and Quantum Optoelectronic Devices, School of Physics, Xi'an Jiaotong University, Xi'an 710049, China}

\author{Chenwei Lv}
\email{lcw205046@gmail.com}
\affiliation{Homer L. Dodge Department of Physics and Astronomy, University of Oklahoma, Norman, OK, 73019, USA}

\author{Ren Zhang}
\email{renzhang@xjtu.edu.cn}
\affiliation{MOE Key Laboratory for Nonequilibrium Synthesis and Modulation of Condensed Matter, Shaanxi Province Key Laboratory of Quantum
Information and Quantum Optoelectronic Devices, School of Physics, Xi'an Jiaotong University, Xi'an 710049, China}
\affiliation{Hefei National Laboratory, Hefei, 230088, China}
\date{\today}

\title{Supplementary Material on ``Geometric Optimization of Quantum Control with Minimum Cost''}

\maketitle

In this Supplemental Material, we provide detailed derivations supporting the results in the main text. Throughout, we set the Planck constant $\hbar = 1$ and the Bohr magneton $\mu_B = 1$ for simplicity. This note is organized into three sections:  
(1) Explicit expressions for the external fields $B_i$ and $\xi_i$ in SU(2) and SU(1,1) systems;  
(2) The metric and geodesic equations on the full parameter manifold;  
(3) The reduced metric and corresponding geodesic equations on the submanifold constrained by fixed state trajectories.

\subsection{1. Explicit expressions of field strength for SU(2) and SU(1,1) symmetric systems}

In this section, we derive the explicit forms of the field strength \( B_i \) for systems with SU(2) dynamical symmetry, and \( \xi_i \) for those with SU(1,1) symmetry.
We begin with the SU(2) case. The evolution operator can be parameterized as
\begin{equation}
\hat{U}(t) = \exp\left(-i\frac{\phi(t)}{2}\hat{\sigma}_z\right) \exp\left(-i\frac{\theta(t)}{2}\hat{\sigma}_y\right) \exp\left(-i\eta(t)\hat{\sigma}_z\right),
\end{equation}
where $\phi(t), \theta(t)$ and $\eta(t)$ are time-dependent parameters. For brevity, we omit the explicit time dependence in the following discussion.
We expand the evolution operator by the power series,
\begin{equation}
\begin{aligned}
\exp(-i\eta\hat{\sigma}_z) &= \mathbb{I} - i\eta \hat{\sigma}_z - \frac{\eta^2}{2!} \mathbb{I} + i\frac{\eta^3}{3!} \hat{\sigma}_z + \frac{\eta^4}{4!} \mathbb{I} - \cdots \\
&= \mathbb{I} \cos\eta - i \hat{\sigma}_z \sin\eta, \\
\exp\left(-i\frac{\theta}{2} \hat{\sigma}_y\right) &= \mathbb{I} \cos\frac{\theta}{2} - i \hat{\sigma}_y \sin\frac{\theta}{2}, \\
\exp\left(-i\frac{\phi}{2} \hat{\sigma}_z\right) &= \mathbb{I} \cos\frac{\phi}{2} - i \hat{\sigma}_z \sin\frac{\phi}{2},
\end{aligned}
\end{equation}
where \( \hat{\sigma}_{x,y,z} \) are the Pauli matrices, and \( \mathbb{I} \) is the \( 2 \times 2 \) identity matrix.

Substituting these into the expression for \( \hat{U} \), we find
\begin{equation}
\begin{aligned}
\label{operatorU}
\hat{U}(t) &= \left(\mathbb{I} \cos\frac{\phi}{2} - i \hat{\sigma}_z \sin\frac{\phi}{2}\right) \left(\mathbb{I} \cos\frac{\theta}{2} - i \hat{\sigma}_y \sin\frac{\theta}{2}\right) \left(\mathbb{I} \cos\eta - i \hat{\sigma}_z \sin\eta\right) \\
&= \cos\frac{\theta}{2} \cos\left(\eta + \frac{\phi}{2}\right)\mathbb{I} - i \sin\frac{\theta}{2} \sin\left(\eta - \frac{\phi}{2}\right) \hat{\sigma}_x \\
&\quad - i \sin\frac{\theta}{2} \cos\left(\eta - \frac{\phi}{2}\right) \hat{\sigma}_y - i \cos\frac{\theta}{2} \sin\left(\eta + \frac{\phi}{2}\right) \hat{\sigma}_z,
\end{aligned}
\end{equation}
where we have used the identity \( \hat{\sigma}_a \hat{\sigma}_b = i \epsilon_{abc} \hat{\sigma}_c \) for distinct \( a \neq b \), and \( \hat{\sigma}_a^2 = \mathbb{I} \).

Acting on the initial state $( 0,1 )^{\rm T}$, the evolution operator yields
\begin{equation}
\hat{U}(t)\begin{pmatrix} 0 \\ 1 \end{pmatrix} = e^{i\eta}
\begin{pmatrix}
- e^{-i\phi/2} \sin\frac{\theta}{2} \\
\;\;\; e^{i\phi/2} \cos\frac{\theta}{2}
\end{pmatrix}.
\end{equation}

The inverse evolution operator is
\begin{equation}
\begin{aligned}
\label{inverseU}
\hat{U}^{-1}(t) &= \left(\mathbb{I} \cos\eta + i \hat{\sigma}_z \sin\eta\right)\left(\mathbb{I} \cos\frac{\theta}{2} + i \hat{\sigma}_y \sin\frac{\theta}{2}\right)\left(\mathbb{I} \cos\frac{\phi}{2} + i \hat{\sigma}_z \sin\frac{\phi}{2}\right) \\
&= \cos\frac{\theta}{2} \cos\left(\eta + \frac{\phi}{2}\right)\mathbb{I} + i \sin\frac{\theta}{2} \sin\left(\eta - \frac{\phi}{2}\right) \hat{\sigma}_x \\
&\quad + i \sin\frac{\theta}{2} \cos\left(\eta - \frac{\phi}{2}\right) \hat{\sigma}_y + i \cos\frac{\theta}{2} \sin\left(\eta + \frac{\phi}{2}\right) \hat{\sigma}_z.
\end{aligned}
\end{equation}

Using Eqs.~\eqref{operatorU} and \eqref{inverseU}, the Hamiltonian can be obtained through 
\[
\hat{H}(t) = i\left(\partial_t \hat{U}(t)\right)\hat{U}^{-1}(t),
\]
which is a linear combination of Pauli matrices. The coefficients correspond to the components of the external magnetic field. Using the identity \( \operatorname{Tr}[\hat{\sigma}_i \hat{\sigma}_j] = 2\delta_{ij} \), the magnetic field components are extracted as
\begin{equation}
\begin{aligned}
B_x(t) &= (\partial_t \eta)\cos\phi \sin\theta - \frac{1}{2}(\partial_t \theta)\sin\phi, \\
B_y(t) &= (\partial_t \eta)\sin\phi \sin\theta + \frac{1}{2}(\partial_t \theta)\cos\phi, \\
B_z(t) &= (\partial_t \eta)\cos\theta + \frac{1}{2}(\partial_t \phi).
\label{Eq_Bi}
\end{aligned}
\end{equation}
Thus, for any prescribed trajectory \( \{\eta(t), \theta(t), \phi(t)\} \), the corresponding control field \( \vec{B}(t) \) can be obtained from Eq.~\eqref{Eq_Bi}.

We now turn to the SU(1,1) case. Analogously, the evolution operator is given by
\begin{equation}
\hat{U}(t) = \exp\left(-i\phi \hat{K}_0\right) \exp\left(-i\rho \hat{K}_2\right) \exp\left(-i2\eta \hat{K}_0\right),
\end{equation}
and each term in the expression above can be expanded as follows
\begin{equation}
\begin{aligned}
\exp(-i2\eta \hat{K}_0) &= \mathbb{I} \cos\eta - i \hat{K}_0 \sin\eta, \\
\exp(-i\rho \hat{K}_2) &= \mathbb{I} \cosh\frac{\rho}{2} - i \hat{K}_2 \sinh\frac{\rho}{2}, \\
\exp(-i\phi \hat{K}_0) &= \mathbb{I} \cos\frac{\phi}{2} - i \hat{K}_0 \sin\frac{\phi}{2},
\end{aligned}
\end{equation}
where we use the non-Hermitian representation of the \( \mathfrak{su}(1,1) \) algebra
\[
\hat{K}_0 = \frac{1}{2}\hat{\sigma}_z, \quad \hat{K}_1 = \frac{i}{2}\hat{\sigma}_x, \quad \hat{K}_2 = \frac{i}{2}\hat{\sigma}_y.
\]
The product yields the explicit expression of the evolution operator,
\begin{equation}
\begin{aligned}
\hat{U}(t) = &\cosh\frac{\rho}{2} \cos\left(\eta + \frac{\phi}{2}\right) \mathbb{I} - i \sinh\frac{\rho}{2} \sin\left(\eta - \frac{\phi}{2}\right) \hat{K}_1 \\
& - i \sinh\frac{\rho}{2} \cos\left(\eta - \frac{\phi}{2}\right) \hat{K}_2 - i \cosh\frac{\rho}{2} \sin\left(\eta + \frac{\phi}{2}\right) \hat{K}_0.
\end{aligned}
\end{equation}
Then, the time-dependent Hamiltonian is obtained via
\[
\hat{H}(t) = i\left(\partial_t \hat{U}(t)\right)\hat{U}^{-1}(t) = \sum_{i=0}^2 \xi_i(t)\hat{K}_i,
\]
from which the field components are extracted using \( \xi_i(t) = 2\,\text{Tr}[\hat{H}(t)\hat{K}_i^\dagger] \),
\begin{equation}
\begin{aligned}
\xi_0(t) &= 2(\partial_t \eta)\cosh\rho + \partial_t \phi, \\
\xi_1(t) &= 2(\partial_t \eta)\cos\phi \sinh\rho - (\partial_t \rho)\sin\phi, \\
\xi_2(t) &= 2(\partial_t \eta)\sin\phi \sinh\rho + (\partial_t \rho)\cos\phi.
\label{Eq_xi}
\end{aligned}
\end{equation}
Hence, given a control trajectory \( \{\eta(t), \rho(t), \phi(t)\} \), the associated control fields \( \{\xi_0(t), \xi_1(t), \xi_2(t)\} \) are determined explicitly via Eq.~\eqref{Eq_xi}.

\subsection{2. Derivation of geometry metric}
Due to the difficulty in experimentally applying external fields in different directions, we incorporate anisotropic coefficients into the cost function. Specifically, the cost function is written as
\begin{equation}
ds^2 = \sum_{i=x,y,z} {\cal I}_i^2 B_i^2 dt^2,
\label{Eq.CEle}
\end{equation}
allowing our method to adapt to different experimental systems by defining corresponding geometries. Substituting Eq.~(\ref{Eq_Bi}) into Eq.~(\ref{Eq.CEle}), the cost function becomes the line element in the parameter space $\{\eta,\theta,\phi\}$,
\begin{align}
ds^2 = g_{\eta\eta}d\eta^2+g_{\theta\theta}d\theta^2+g_{\phi\phi}d\phi^2+2g_{\eta\theta} d\eta d\theta +2g_{\eta\phi} d\eta d\phi,
\label{Eq.Gms}
\end{align}
which defines a metric on this parameter space. The metric components in Eq.~(\ref{Eq.Gms}) are given by
\begin{equation}
\begin{aligned}
 g_{\eta\eta}& = {\cal I}_z^2\cos^2\theta+{\cal I}_x^2\cos^2\phi\sin^2\theta+{\cal I}_y\sin^2\theta\sin^2\phi,\\
 g_{\theta\theta} & = \frac14({\cal I}_y^2\cos^2\phi+{\cal I}_x^2\sin^2\phi),\\
 g_{\phi\phi}& = \frac14{\cal I}_z^2,\\
 g_{\eta\theta} & =\frac12({\cal I}_y^2-{\cal I}_x^2)\cos\phi\sin\phi\sin\theta, \\
g_{\eta\phi}& = \frac12{\cal I}_z^2\cos\theta.
\end{aligned}
\end{equation}

Once the metric is established, the optimal path in parameter space can be determined by solving the geodesic equations, 
\begin{align}
\frac{d^2\lambda^\mu}{dt^2}+\Gamma_{\nu n}^\mu \frac{d\lambda^\nu}{dt}\frac{d\lambda^n}{dt}=0,
\label{Eq.geo}
\end{align}
where $\lambda^{\mu} \in \{\eta,\theta,\phi\}$, and $\Gamma_{\nu n}^\mu$ denotes the Christoffel symbol which is defined as
\begin{align}
\Gamma_{\nu n}^\mu=\frac12g^{\mu m}\left(\partial_ng_{m\nu}+\partial_\nu g_{mn}-\partial_mg_{\nu n}\right),
\label{Eq.CS}
\end{align}
with $g^{\mu m} = (g^{-1})_{\mu m}$ and $\{\mu,\nu,m,n\}$ being indices in parameter space. In principle, for all SU(2) systems, we can construct distinct geometries tailored to specific experimental scenarios and derive the corresponding optimal trajectories. 

As a concrete example, we consider the isotropic case (${\cal I}_i = 1$), where the metric becomes
\begin{equation}
ds^2 =\left(d\eta+\frac{1}{2}\cos\theta d\phi\right)^2+\frac{1}{4}(d\theta^2+\sin^2\theta d\phi^2),
\label{Eq.Gmi}
\end{equation}
and the corresponding geodesic equations from Eq.~(\ref{Eq.geo}) are as follows,
\begin{equation}
\begin{aligned}
\partial_t^2\eta &= -\cot \theta \, \partial_t\eta \, \partial_t\theta + \frac{1}{2}\csc\theta \, (\partial_t\phi \, \partial_t\theta),\\
\partial_t^2\theta &= -2\sin\theta \, (\partial_t\eta \, \partial_t\phi),\\
\partial_t^2\phi &= 2\csc\theta \, (\partial_t\eta \, \partial_t\theta) - \cot\theta \, (\partial_t\phi \, \partial_t\theta).
\end{aligned}
\label{Eq.su2G}
\end{equation}
Given boundary conditions $(\theta_{\rm i}, \phi_{\rm i}, \eta_{\rm i})$ and $(\theta_{\rm f}, \phi_{\rm f}, \eta_{\rm f})$, solving Eq.~(\ref{Eq.su2G}) yields the optimal trajectory $(\theta(t), \phi(t), \eta(t))$.

Similarly, for a system with SU(1,1) dynamical symmetry, the cost function is
\begin{equation}
ds^2 = \frac14 \sum_{i=0,1,2} {\cal I}_i^2 \xi_i^2 dt^2.
\label{Eq.CEle2}
\end{equation}
Substituting Eq.~(\ref{Eq_xi}) into Eq.~(\ref{Eq.CEle2}), the cost function becomes
\begin{align}
 ds^2 =  (g_{\eta\eta}d\eta^2+2g_{\eta\rho} d\eta d\rho +2g_{\eta\phi} d\eta d\phi)+g_{\rho\rho}d\rho^2+g_{\phi\phi}d\phi^2,
\label{Eq.Gm11}
\end{align}
which defines the metric in the parameter space $\{\eta,\rho,\phi\}$. The metric components are
\begin{equation}
\begin{aligned}
 g_{\eta\eta}= &{\cal I}_0^2\cosh^2\rho+{\cal I}_1^2\cos^2\phi\sinh^2\rho+{\cal I}_2\sinh^2\rho\sin^2\phi,\\
 g_{\rho\rho}= &\frac14({\cal I}_2^2\cos^2\phi+{\cal I}_1^2\sin^2\phi)d\rho^2 ,\\ 
 g_{\phi\phi}=&\frac14{\cal I}_0^2,\\
 g_{\eta\rho}=& \frac12({\cal I}_2^2-{\cal I}_1^2)\cos\phi\sin\phi\sinh\rho ,\\
g_{\eta\phi}=& \frac12{\cal I}_0^2\cosh\rho. 
\end{aligned}
\end{equation}
For the isotropic case (${\cal I}_{i=0,1,2}=1$), the geodesic equations can  be written as
\begin{equation}
\begin{aligned}
\partial_t^2\eta &= -3\coth\rho\, (\partial_t\eta\, \partial_t\rho) - \frac{1}{2}\csch\rho\, (\partial_t\rho\, \partial_t\phi), \\
\partial_t^2\rho &= 4\sinh(2\rho)(\partial_t\eta)^2 + 2\sinh\rho\, (\partial_t\eta\, \partial_t\phi), \\
\partial_t^2\phi &= \frac{4 + 2\cosh(2\rho)}{\sinh\rho}(\partial_t\eta\, \partial_t\rho) + \coth\rho\, (\partial_t\rho\, \partial_t\phi).
\end{aligned}
\end{equation}

If a trajectory $\{\theta(t), \phi(t)\}$ (SU(2)) or $\{\rho(t), \phi(t)\}$ (SU(1,1)) is prescribed, we can choose a proper $\eta(t)$ to minimize its contribution to the cost. For SU(2) systems, the cost function can be recast as
\begin{align}
ds^2 = (g_{\eta\eta}d\eta^2+2g_{\eta\theta} d\eta d\theta +2g_{\eta\phi} d\eta d\phi)+g_{\theta\theta}d\theta^2+g_{\phi\phi}d\phi^2,
\end{align}
and the optimal $\eta_{\rm opt}(t)$ satisfies
\begin{align}
d\eta = -\frac{g_{\eta\theta}d\theta+g_{\eta\phi}d\phi}{g_{\eta\eta}}.
\end{align}
Then the optimal phase $\eta_{\rm opt}(t)$ is obtained by 
\begin{equation}
\eta_{\rm opt}(t) = -\int_{t_i}^{t}\frac{g_{\eta\theta}\partial_\tau\theta(\tau)+g_{\eta\phi}\partial_\tau\phi(\tau)}{g_{\eta\eta}}d\tau,
\label{Eq.Oeta2}
\end{equation}
which yields the optimal Hamiltonian along $\{\theta(t), \phi(t)\}$ from Eq.~(\ref{Eq_Bi}). In the next section, we will show that this optimal phase corresponds to a geodesic on the submanifold.

For SU(1,1) systems, the cost function can be rewritten as
\begin{align}
ds^2 = (g_{\eta\eta}d\eta^2+2g_{\eta\rho} d\eta d\rho +2g_{\eta\phi} d\eta d\phi)+g_{\rho\rho}d\rho^2+g_{\phi\phi}d\phi^2,
\end{align}
and the optimal phase satisfies
\begin{align}
d\eta = -\frac{g_{\eta\rho}d\rho+g_{\eta\phi}d\phi}{g_{\eta\eta}}.
\end{align}
Similarly, we obtain the optimal phase $\eta_{\rm opt}(t)$ by
\begin{equation}
\eta_{\rm opt}(t) = -\int_{t_i}^{t}\frac{g_{\eta\rho}\partial_\tau\rho(\tau)+g_{\eta\phi}\partial_\tau\phi(\tau)}{g_{\eta\eta}}d\tau,
\label{Eq.Oeta11}
\end{equation}
which also yields the optimal Hamiltonian along $\{\rho(t), \phi(t)\}$ from Eq.~(\ref{Eq_xi}).

To conclude, our scheme constructs feasible geometric manifolds tailored to specific experimental setups, such that geodesics on these manifolds represent optimal evolution trajectories. Furthermore, given a desired trajectory $\{\theta(t),\phi(t)\}$ or $\{\rho(t),\phi(t)\}$, the method provides an optimal phase $\eta_{\rm opt}(t)$ and the corresponding driving fields $\vec{B}(t)$, facilitating practical experimental realization even in anisotropic conditions.

\subsection{3. The reduced metric and geodesic equations for the submanifold}
\begin{figure}[t]
	\centering
	\includegraphics[width=0.7\textwidth]{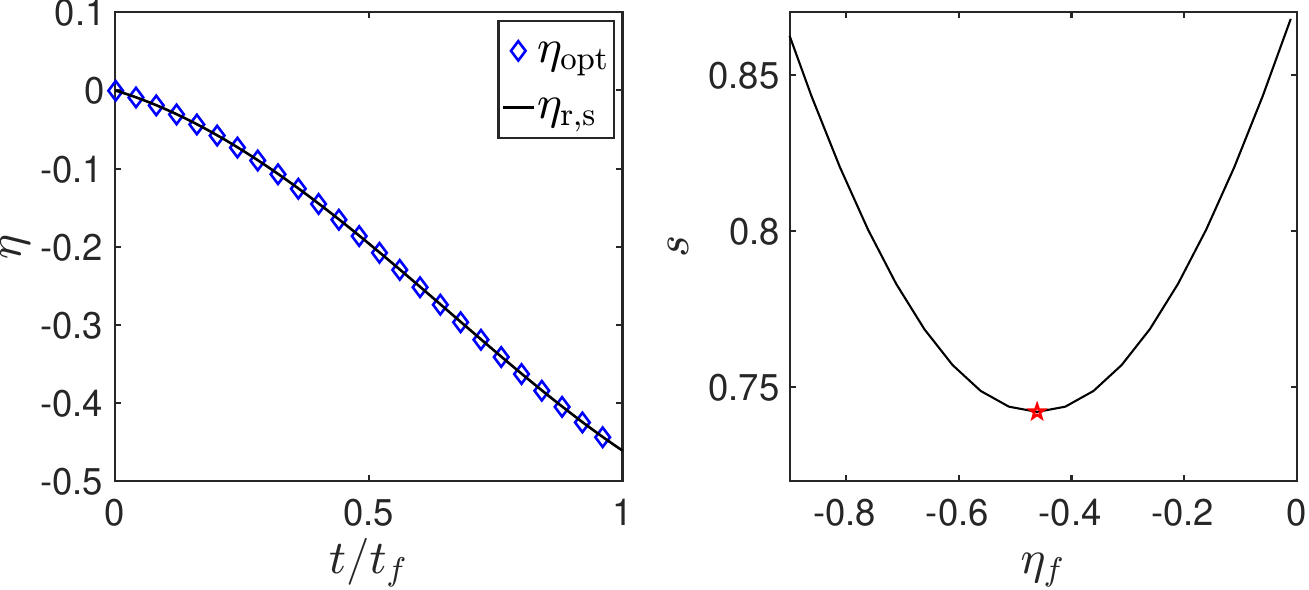}
	\caption{ Comparison of optimal phase from integration and subspace geodesics.
Left: The optimal phase from the integration in Eq.~(\ref{etaoptsu2}) (blue diamonds) agrees precisely with that obtained from the geodesic (black solid curve), validating the geometric approach.
Right: Total arc length of geodesics in parameter space as a function of the final phase (black curve). The red pentagram indicates the optimal phase $\eta_f^* = -0.461$,  corresponding to the global minimum arc length {$s(\eta_f^*)$}.}
	\label{Figrm}
\end{figure}

In this section, we illustrate that the trajectory with the optimal $\eta_{\rm opt}(t)$ obtained from Eq.~(\ref{Eq.Oeta2}) and (\ref{Eq.Oeta11}) corresponds to the geodesic on the submanifold spanned by $(t,\eta)$. As in previous sections, we first consider the SU(2) case. For simplicity, we examine an isotropic system with ${\cal I}_i = 1$ and choose a specific trajectory denoted by
\begin{align}
\theta(t) = \frac{t}{t_f} + \frac{1}{5},\ \phi(t) = \theta^2(t),
\label{Eq.phi}
\end{align}
Substituting Eq.~(\ref{Eq.phi}) into Eq.~(\ref{Eq.Gmi}), we obtain the reduced metric
\begin{align}
ds^2 = d\eta^2 + 2\theta\cos\theta \, d\eta \, d\theta + \frac{1}{4}\left(1+4\theta^2\right) \, d\theta^2.
\end{align}
Since $\theta$ and $t$ are linearly related, the submanifold $\{\theta,\eta\}$ is directly related to the manifold $\{t,\eta\}$. Therefore, we could solve the geodesic equations in $\{\theta,\eta\}$ and extract the minimum cost. The nonzero Christoffel symbols for this metric are written as
\begin{equation}
\begin{aligned}
\Gamma_{\theta\theta}^\eta &= \frac{-\cos\theta+\theta(1+4\theta^2)\sin \theta}{1+4\theta^2\sin^2\theta},\\
\Gamma_{\theta\theta}^\theta &= \frac{4\theta(1+\theta\cot\theta)}{\csc^2\theta+4\theta^2}.
\end{aligned}
\end{equation}
Solving the geodesic equations Eq.~(\ref{Eq.geo}) with boundary conditions $\{0.2, 0\}$ and $\{1.2, \eta_f\}$ yields the geodesic trajectory $\eta_{\rm r}(t)$ in the submanifold. By properly selecting $\eta_f$, we identify the shortest geodesic, denoted $\eta_{\rm r,s}(t)$, which precisely matches the result obtained from direct integration
\begin{equation}
\label{etaoptsu2}
\eta_{\rm opt}(t) = -\frac{1}{2} \int_{t_{\rm i}}^{t} \cos\theta(\tau) \left[\partial_\tau\phi(\tau)\right] d\tau,
\end{equation}
which is the explicit form of Eq.~(\ref{Eq.Oeta2}).
As shown in the left panel of Fig.~\ref{Figrm},  the black solid curve represents the phase $\eta_{\rm r,s}(t)$ along the geodesic in the submanifold, while blue diamonds denote the phase $\eta_{\rm opt}(t)$ obtained from Eq.~(\ref{etaoptsu2}). Agreement between the two methods is observed. The right panel of Fig.~\ref{Figrm} shows the total arc lengths of geodesics for selected final phases (black curve). The red pentagram marks the phase chosen in the left panel of Fig.~\ref{Figrm}, which exhibits the minimal arc length, confirming the reliability of our geometric approach.

\begin{figure}[ht]
	\centering
	\includegraphics[width=0.7\textwidth]{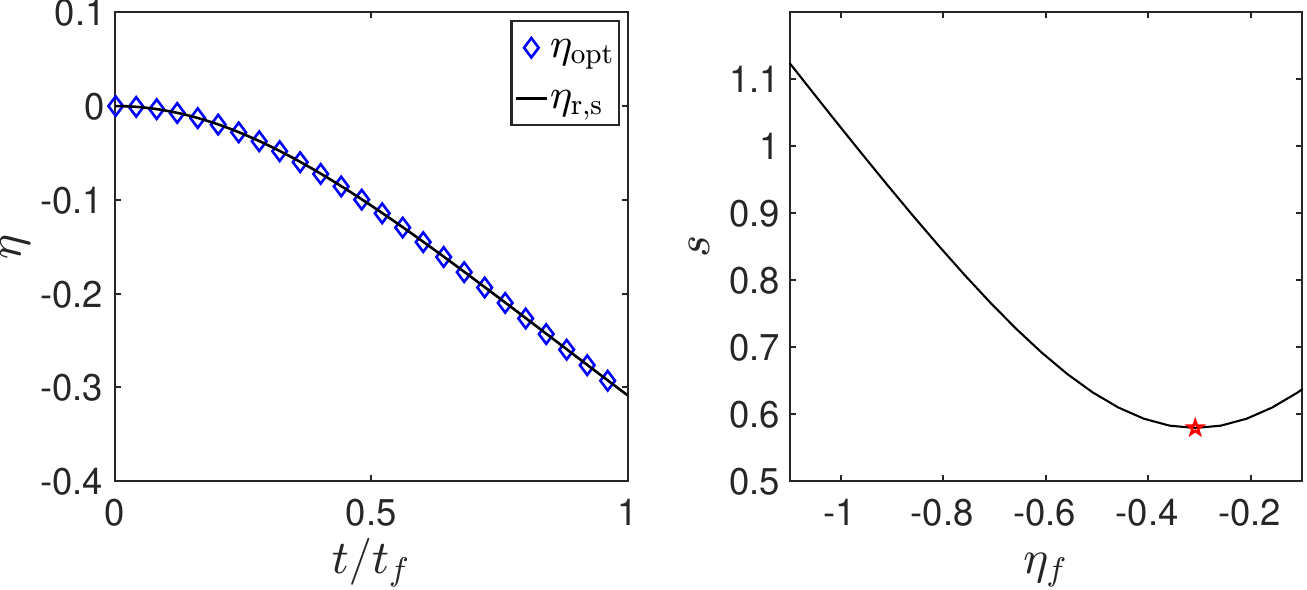}
	\caption{Comparison of optimal phase from integration and submanifold geodesics for SU(1,1).
Left: The integrated optimal phase (blue diamonds) agrees with the geodesic-based result (black solid curve), validating our geometric approach. 
Right: Geodesic arc length versus final phase (black curve). The red pentagram marks the optimal $\eta_f = -0.309$, confirming global minimality.}
	\label{Figrm11}
\end{figure}

For a system with SU(1,1) dynamical symmetry, the optimal $\eta(t)$ derived from Eq.~(\ref{Eq.Oeta11}) similarly corresponds to the geodesic in the $(t,\eta)$ subspace. To demonstrate this, we again consider the isotropic case with ${\cal I}_{i=0,1,2} = 1$ and a representative trajectory
\begin{equation}
\begin{aligned}
\rho(t) &= \frac{t}{t_f} + \frac{1}{5},\ \phi(t) = \rho^2(t).
\end{aligned}
\end{equation}
Under this trajectory, the metric in Eq.~(\ref{Eq.Gm11}) simplifies to
\begin{align}
ds^2 = \cosh(2\rho)\, d\eta^2 + 2\rho\cosh(\rho)\, d\eta \, d\rho + \left(\frac14+\rho^2\right)\, d\rho^2.
\end{align}
The corresponding Christoffel symbols include six nonzero components,
\begin{equation}
\begin{aligned}
 \Gamma_{\eta\eta}^\eta &= \frac{8\rho\cosh^2\rho\sinh\rho}{\cosh(2\rho)(1+2\rho^2)-2\rho^2},\\
 \Gamma_{\eta\rho}^\eta &= \frac{(1+4\rho^2)\sinh(2\rho)}{\cosh(2\rho)(1+2\rho^2)-2\rho^2},\\
 \Gamma_{\rho\rho}^\eta &= \frac{\cosh\rho+(1+4\rho^2)\rho\sinh(\rho)}{\cosh(2\rho)(1+2\rho^2)-2\rho^2},\\
 \Gamma_{\eta\eta}^\rho &= -\frac{2\sinh(4\rho)}{\cosh(2\rho)(1+2\rho^2)-2\rho^2},\\
 \Gamma_{\eta\rho}^\rho &= -\frac{4\rho\cosh\rho\sinh(2\rho)}{\cosh(2\rho)(1+2\rho^2)-2\rho^2},\\
 \Gamma_{\rho\rho}^\rho &= \frac{4\rho\sinh(\rho)(\sinh\rho-\rho\cosh\rho)}{\cosh(2\rho)(1+2\rho^2)-2\rho^2}.
\end{aligned}
\end{equation}
Solving Eq.~(\ref{Eq.geo}) under the boundary conditions $(0.2, 0)$ and $(1.2, \eta_f)$, we obtain the geodesic in the $\{\rho, \eta\}$ submanifold. As shown in the left panel of Fig.~\ref{Figrm11}, the optimal phase from direct integration of Eq.~(\ref{Eq.Oeta11}) (blue diamonds) matches exactly with the geodesic result (black solid curve). The optimal final phase $\eta_f = -0.309$, indicated by the red pentagram in the right panel, corresponds to the globally minimal arc length in the reduced subspace.

\bibliographystyle{apstest}
\bibliography{ref_SM.bib}